\documentclass{iopart}
\usepackage{iopams,mathptm,times,graphicx}

\begin{document}
\title[Integrable discretizations for the short wave model]
{Integrable discretizations for the short wave model of the 
Camassa-Holm equation}
\author{Bao-Feng Feng$^{1}\footnote{e-mail: feng@utpa.edu}$, 
Ken-ichi Maruno
$^{1}\footnote{e-mail: kmaruno@utpa.edu}$ and 
Yasuhiro Ohta$^{2}$ 
}
\address{$^1$~Department of Mathematics,
The University of Texas-Pan American,
Edinburg, TX 78541
}
\address{$^2$~Department of Mathematics,
Kobe University, Rokko, Kobe 657-8501, Japan
}
\date{\today}
\def\submitto#1{\vspace{28pt plus 10pt minus 18pt}
     \noindent{\small\rm To be submitted to : {\it #1}\par}}

\begin{abstract}
The link between the short wave model of the Camassa-Holm equation (SCHE)
and bilinear equations of the two-dimensional Toda lattice (2DTL)
is clarified. The parametric form of
$N$-cuspon solution of the SCHE in Casorati determinant is then
given. Based on the above finding, integrable semi-discrete and
full-discrete analogues of the SCHE are constructed. The
determinant solutions of both semi-discrete and fully discrete
analogues of the SCHE are also presented.
\par
\kern\bigskipamount\noindent
\today
\end{abstract}

\kern-\bigskipamount

\submitto{\JPA}

\section{Introduction}
In the present paper, we consider integrable discretizations of the
nonlinear partial differential equation
\begin{equation}
w_{TXX}-2\kappa^2w_X+2w_Xw_{XX}+ww_{XXX}= 0,\label{HS-eq}
\end{equation}
which belongs to the Harry-Dym hierarchy \cite{Kruskal,ACHM,ACFHM}.
Here $\kappa$ is a real parameter and, as shown subsequently, can be
normalized by the scaling transformation when $\kappa \ne 0$. A
connection between Eq.(\ref{HS-eq}) and the sinh-Gordon equation
was established in \cite{HHD}. When $\kappa = 0$, Eq.(\ref{HS-eq})
is called the Hunter-Saxton equation and is derived as a model for weakly
nonlinear orientation waves in massive nematic liquid crystals
\cite{HS_SIAM}. The Lax pair and bi-Hamiltonian structure were
discussed by Hunter and Zheng \cite{HS_PhysD}. The dissipative and
dispersive weak solutions were discussed in details by the same
authors \cite{HS_ARMA1,HS_ARMA2}.

Equation (\ref{HS-eq}) can be viewed as a short-wave model of the
Camassa-Holm equation \cite{CH}
\begin{equation}
w_T+2\kappa^2w_X-w_{TXX}+3ww_X=2w_Xw_{XX}+ww_{XXX}\,.\label{CH-eq}
\end{equation}
Following the procedure in \cite{Manna1,Manna2,Manna3}, we introduce the time and
space variables ${\tilde T}$ and ${\tilde X}$ 
$$
{\tilde T}= \epsilon T\,,\quad  {\tilde X}=\epsilon^{-1} X\,,
$$
where $\epsilon$ is a small parameter.
Then $w$ is expanded as $w= \epsilon^2 (w_0+\epsilon w_1+\cdots)$
with $w_i$ ($i=0,1,\cdots$) being functions of ${\tilde T}$ and ${\tilde X}$.
At the lowest order in $\epsilon$, we obtain
\begin{equation}
w_{0,{\tilde T}{\tilde X}{\tilde X}}-2\kappa^2w_{0,{\tilde X}}
+2w_{0,{\tilde X}}w_{0,{\tilde X}{\tilde X}}+w_{0}w_{0,{\tilde X}{\tilde X}{\tilde X}}= 0\,,
\end{equation}
which is exactly Eq.(\ref{HS-eq}) after writing back into the original variables.
Based on this fact, Matsuno obtained the
$N$-cuspon solution of Eq.(\ref{HS-eq}) by taking the short-wave limit
on the $N$-soliton solution of
the Camassa-Holm equation \cite{MatsunoPLA,Matsuno}.

Note that the parameter $\kappa$ of Eq.(\ref{HS-eq}) can be normalized to $1$  under the transformation
\[
x= \kappa X\,,\quad   t= \kappa T\,,
\]
which leads to
\begin{equation}
w_{txx}-2 w_x+ 2w_xw_{xx}+ww_{xxx} = 0\,.\label{HS-eq2}
\end{equation}
We call Eq.({\ref{HS-eq2}}) the short wave model of the Camassa-Holm
equation (SCHE). Without loss of generality, we will focus on Eq.
(\ref{HS-eq2}) and its integrable discretizations, since the
solution of Eq.(\ref{HS-eq}) with arbitrary nonzero $\kappa$, its
integrable discretizations and the corresponding solutions can be
recovered through the above transformation.

The reminder of the present paper is organized as follows. In
section 2, we reveal a connection between the SCHE and the bilinear form 
two-dimensional Toda-lattice (2DTL) equations.
The parametric form of $N$-cuspon solution expressed by the
Casorti determinant is given, which is consistent with the solution
given in \cite{MatsunoPLA}. Based on this fact, we propose an integrable 
semi-discrete analogue of the SCHE in section 3, and further its integrable
full-discrete analogue in section 4. The concluding remark is given in section 5.

\section{The connection with 2DTL equations, and $N$-cuspon solution in determinant form}
\subsection{The link of the SCHE with the two-reduction of 2DTL equations}
In this section, we will show that the SCHE can be derived
from the bilinear form of two-dimensional Toda lattice (2DTL)
equations
\begin{equation}\label{2DTD}
  -\left(\frac{1}{2}D_{-1}D_{1}-1 \right) \tau_n \cdot \tau_n = \tau_{n+1} \tau_{n-1}\,,
\end{equation}
where $D_x$ is the Hirota $D$-derivative defined as
\[
D_x^nf\cdot g=\left(\frac{\partial}{\partial x}
-\frac{\partial}{\partial y}\right)^nf(x)g(y)|_{y=x}\,,
\]
and $D_{-1}$ and $D_{1}$ represent the Hirota $D$ derivatives with
respect to variables $x_{-1}$ and $x_1$, respectively.

It is shown that the $N$-soliton solution of the 2DTL equations
(\ref{2DTD}) can be expressed as the Casorati determinant
\cite{2Dtoda,Hirota_book}
\begin{equation}\label{2DTL_solution}
   \tau_n=\left|
\psi_{i}^{(n+j-1)}(x_{1},x_{-1})
\right|_{1\leq i,j\leq N}
=\left|\matrix{
\psi_1^{(n)} &\psi_1^{(n+1)} &\cdots &\psi_1^{(n+N-1)} \cr
\psi_2^{(n)} &\psi_2^{(n+1)} &\cdots &\psi_2^{(n+N-1)} \cr
\vdots       &\vdots         & \ddots      &\vdots           \cr
\psi_N^{(n)} &\psi_N^{(n+1)} &\cdots &\psi_N^{(n+N-1)}}\right|\,,
\end{equation}
with $\psi_i^{(n)}$ satisfying the following dispersion relations:
$$
 \frac{\partial \psi_i^{(n)}}{\partial x_{-1}}  = \psi_i^{(n-1)}, \qquad
  \frac{\partial \psi_i^{(n)}}{\partial x_{1}} = \psi_i^{(n+1)}\,.
$$
A particular choice of $\psi_i^{(n)}$
\begin{equation}\label{phiin}
\psi_i^{(n)}=a_{i,1}p_i^ne^{{p_i}^{-1} x_{-1} +p_ix_1+\eta_{0i}}
+a_{i,2}q_i^ne^{{q_i}^{-1} x_{-1}+q_ix_1+\eta'_{0i}}\,,
\end{equation}
automatically satisfies the
above dispersion relations.

Applying the two-reduction
$\tau_{n-1}=(\prod_{i=1}^Np^2_i)^{-1} \tau_{n+1}$, i.e., enforcing $p_i=-q_i$, $i=1, \cdots, N$,
we get
\begin{equation}\label{2DTL_2reduction}
  -\left(\frac{1}{2}D_{-1}D_{1}-1 \right) \tau_n \cdot \tau_n = \tau^2_{n+1}\,,
\end{equation}
where the gauge transformation $\tau_n \to (\prod_{i=1}^Np_i)^{n} \tau_n$ is used.
Letting $\tau_0=f$, $\tau_1=g$ and $x_{-1}=s$, $x_1=y$, the above bilinear
equation (\ref{2DTL_2reduction}) takes the following form:
\begin{eqnarray}
  -\left(\frac{1}{2}D_sD_y-1\right)f\cdot f =g^2\,, \label{2DTD1} \\
 -\left(\frac{1}{2}D_sD_y-1\right)g\cdot g =f^2\,. \label{2DTD2}
\end{eqnarray}
Introducing $u={g}/{f}$, Eqs.(\ref{2DTD1}) and (\ref{2DTD2}) can be
converted into
\begin{eqnarray}
   -(\ln f)_{ys}+1 &=& u^2\,, \label{log1} \\
-(\ln g)_{ys}+1 &=& u^{-2}\,. \label{log2}
\end{eqnarray}
Subtracting Eq.(\ref{log2}) from Eq.(\ref{log1}), one obtains
\begin{equation}\label{2DTD2e}
\frac{\rho}{2} (\ln \rho)_{ys}+1 = \rho^2\,,
\end{equation}
by letting $\rho = u^2$.

Introducing the dependent variable transformation
$$
w=-2 (\ln g)_{ss}\,,
$$
it then follows
$$
\frac 12 w_y = -\frac{\rho_s}{\rho^2}\,,
$$
or
\begin{equation}\label{BL1}
(\ln \rho)_s = -\frac {\rho}{2} w_y\,,
\end{equation}
by differentiating Eq.(\ref{log2}) with respect to $s$.

In view of Eq.(\ref{BL1}), Eq.(\ref{2DTD2e}) becomes
\begin{equation}\label{BL2}
- \frac{\rho}{2} \left(\frac{\rho}{2} w_y\right)_{y}+1 = \rho^2\,.
\end{equation}
Introducing the hodograph transformation
$$
\left\{\begin{array}{l}
x=2y- 2(\ln g)_s\,,
\\
t=s\,,
\end{array}\right.
$$
and referring to Eq.(\ref{log2}), we have
$$
\frac{\partial x}{\partial y}=2-2 \left(\ln g\right)_{ys} = \frac{2}{\rho}\,,
\qquad
\frac{\partial x}{\partial s}=-2 (\ln g)_{ss}=w\,,
$$
which implies
$$
\left\{\begin{array}{l}\displaystyle
\partial_y=\frac{2}{\rho}\partial_x\,,
\\[5pt]
\partial_s=\partial_t+w\partial_x\,.
\end{array}\right.
$$
Thus, Eqs.(\ref{BL1}) and (\ref{BL2}) can be cast into
\begin{equation}
\left\{\begin{array}{l}\displaystyle
(\partial_t+w\partial_x)\ln\rho=-w_x\,,
\\[5pt]\displaystyle
-w_{xx}+1=\rho^2\,.
\end{array}\right.\label{ch-derive}
\end{equation}
By eliminating $\rho$, we arrive at
$$
(\partial_t+w\partial_x)\ln\left(-w_{xx}+ 1\right)=-2w_x\,,
$$
or
$$
(\partial_t+w\partial_x)w_{xx}
-2w_x\left(1-w_{xx}\right)=0\,,
$$
which is actually the SCHE (\ref{HS-eq2}).
\subsection{The $N$-cuspon solution of the SCHE}

Based on the link of the SCHE with the two-reduction of 2DTL equations, 
the $N$-cuspon solution of
the SCHE (\ref{HS-eq2}) is given as follows:
$$
w=-2 (\ln g)_{ss}\,,
$$
$$
\left\{\begin{array}{l} x=2y- 2(\ln g)_s\,,
\\
t=s\,,
\end{array}\right.
$$
$$
   g=\left|
\psi_{i}^{(j)}(y,s) \right|_{1\leq i,j\leq N}\,\,,
$$
\begin{equation}
\psi_i^{(j)}=a_{i,1}p_i^je^{{p_i}^{-1} s +p_iy+\eta_{0i}}
+a_{i,2}(-p_i)^je^{-{p_i}^{-1}s-p_iy+\eta'_{0i}}\,.
\end{equation}

\begin{figure}[htbp]
\centerline{
\includegraphics[scale=0.43]{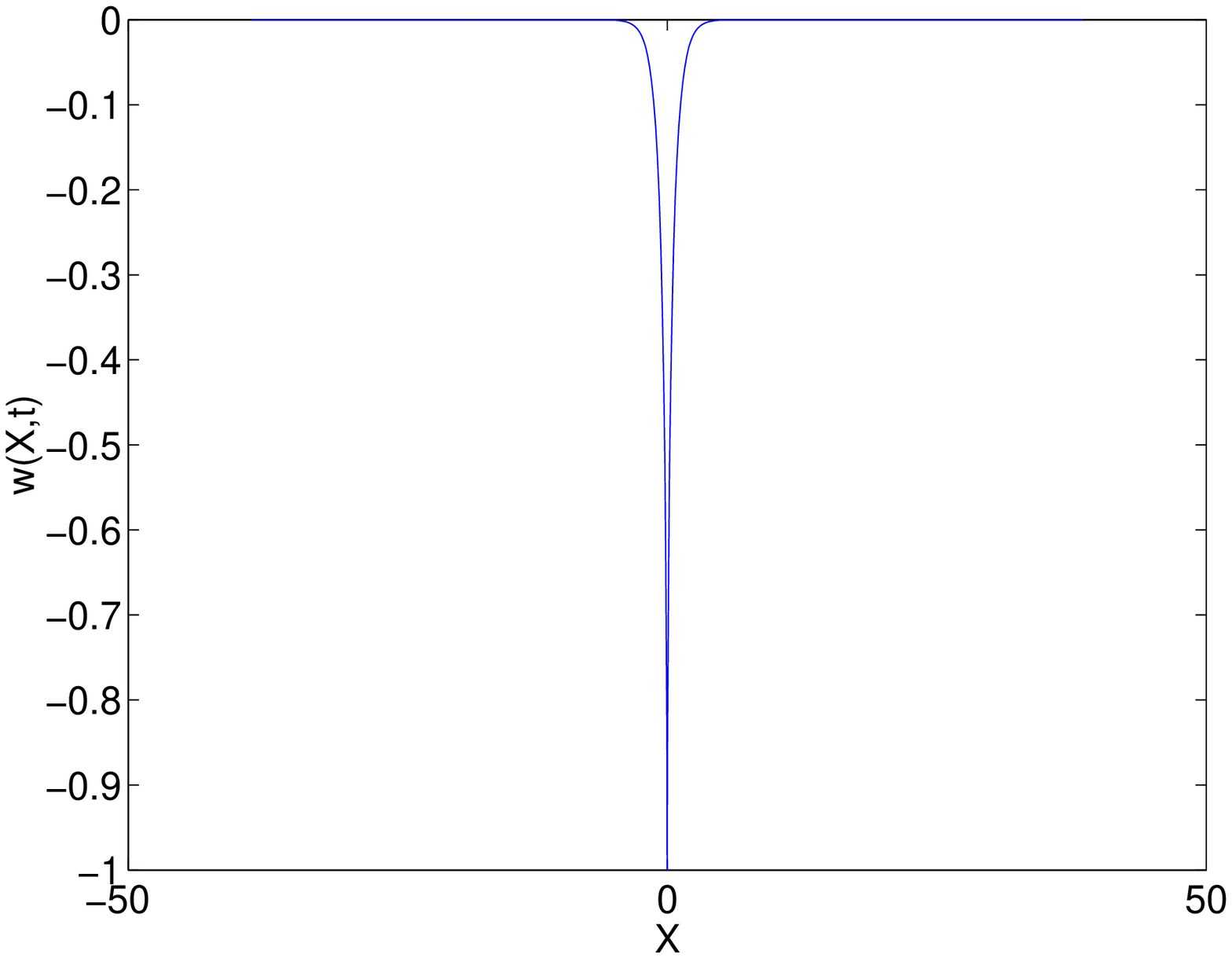}\quad
\includegraphics[scale=0.43]{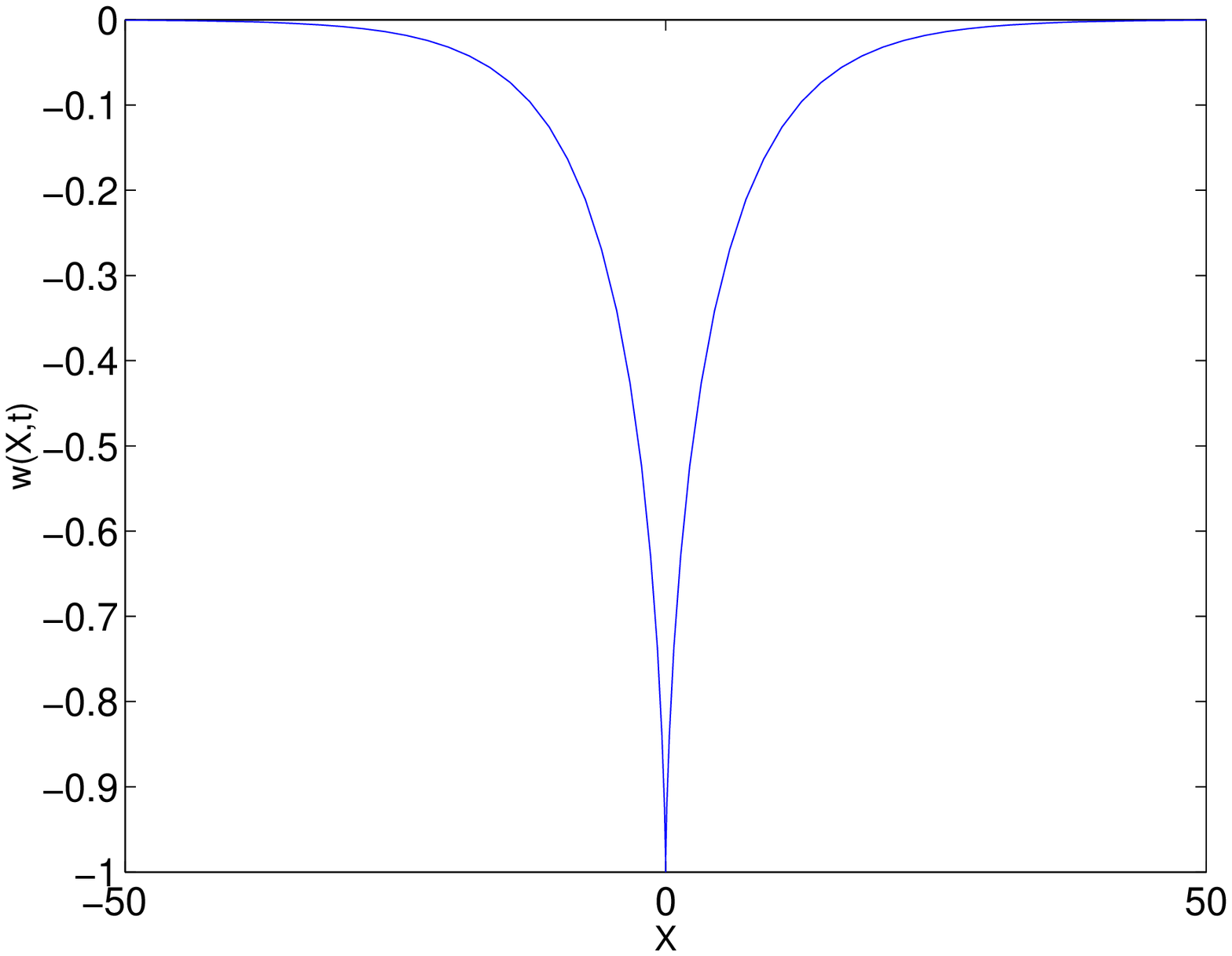}}
\kern-0.4\textwidth \hbox to
\textwidth{\hss(a)\kern19em\hss(b)\kern12em}
\kern+0.4\textwidth
\caption{Plots for
one-cuspon solution for $p_1=\sqrt{2}$ and different $\kappa$:
(a) $\kappa=1.0$; (b) $\kappa=0.1$.}
 \label{f:1cuson}
\end{figure}

Moreover, the $N$-cuspon solution of the SCHE (\ref{HS-eq}) with non-zero $\kappa$ is given as follows:
\begin{equation}\label{N-cuspon-w}
    w(y,T)=-2(\ln g)_{ss},
\end{equation}
\begin{equation}\label{N-cuspon-X}
\left\{\begin{array}{l}
X=\frac{2y}{\kappa}-\frac{2}{\kappa} (\ln g)_s,
\\
T=\frac{s}{\kappa}\,,
\end{array}\right.
\end{equation}
where
$$
g=\left|
\psi_{i}^{(j)}(y,s)
\right|_{1\leq i,j\leq N}\,,
$$
with
$$
\psi_i^{(n)}=a_{i,1}p_i^ne^{p_iy+s/p_i+\eta_{i0}}+a_{i,2}
(-p_i)^n e^{-p_i y - s/p_i + \eta'_{i0}}\,.
$$
We remark here that to assure
the regularity of the solution, the $\tau$-function
is required to be positive definite. In what follows, we list
the one-cuspon and
two-cuspon solutions. For $N=1$, the $\tau$-function is
$$
g = 1+ e^{2p_1(y+\kappa T/p^2_1+y_0)}\,,
$$
by choosing $a_{1,1}/a_{1,2}=-1$, which yields the one-cuspon solution
$$
w(y,T) = -\frac{2}{p^2_1} {\mbox{sech}}^2 \left[p_1(y+\kappa T/p^2_1+y_0)\right]\,,
$$
$$
X=\frac{2y}{\kappa}-\frac{2}{\kappa p_1} \left\{1+\tanh \left[p_1(y+\kappa T/p^2_1+y_0)\right]\right\}\,.
$$
The profiles of one-cuspon  with $\kappa=1.0$ and $\kappa=0.1$ are plotted in Fig. \ref{f:1cuson}.

The $\tau$-function corresponding to the two-cuspon solution is
$$
g = 1+ e^{\theta_1}+e^{\theta_2}+\left(\frac{p_1-p_2}{p_1-p_2}
\right)^2 e^{\theta_1+\theta_2}\,,
$$
with
$$
\theta_i=2p_i(y+\kappa T/p^2_i+y_{i0}), \ i=1,2\,.
$$
Here $a_{1,1}/a_{1,2}=-1$ and $a_{2,1}/a_{2,2}=1$ are chosen to assure
the regularity of the solution.
\section{Integrable semi-discretization of the SCHE}
Based on the link of the SCHE with the two-reduction of 2DTL equations clarified
in the previous section, we attempt to construct the integrable
semi-discrete analogue of the SCHE.

Consider a Casorati determinant
$$
\tau_n(k)=\left|
\psi_{i}^{(n+j-1)}(k)
\right|_{1\leq i,j\leq N}
=\left|\matrix{
\psi_1^{(n)}(k) &\psi_1^{(n+1)}(k) &\cdots &\psi_1^{(n+N-1)}(k) \cr
\psi_2^{(n)}(k) &\psi_2^{(n+1)}(k) &\cdots &\psi_2^{(n+N-1)}(k) \cr
\vdots            &\vdots              &  \ddots     &\vdots                \cr
\psi_N^{(n)}(k) &\psi_N^{(n+1)}(k) &\cdots
 &\psi_N^{(n+N-1)}(k)}\right|
\,,
$$
with $\psi_i^{(n)}$ satisfies the following dispersion relations
\begin{eqnarray}
&&\Delta_k\psi_i^{(n)} = \psi_i^{(n+1)}\,,
\label{k-dispersion-d}\\
&&
\partial_s\psi_i^{(n)} = \psi_i^{(n-1)}\,,
\label{t-linear-d}
\end{eqnarray}
where $\Delta_k$ is
defined as $\Delta_k \psi(k)=\frac{\psi(k)-\psi(k-1)}{a}$.
In particular, we can choose $\psi_i^{(n)}$ as
$$
\psi_i^{(n)}(k)=p_i^n(1-ap_i)^{-k}e^{\xi_i}
+q_i^n(1-aq_i)^{-k}e^{\eta_i}\,,
$$
$$
\xi_i=\frac{1}{p_i}s+\xi_{i0}\,,\quad
\eta_i=\frac{1}{q_i}s+\eta_{i0}\,,
$$
which automatically satisfies the dispersion relations
(\ref{k-dispersion-d}) and (\ref{t-linear-d}). The above Casorati
determinant satisfies the bilinear form of the semi-discrete 2DTL
equation (the B\"acklund transformation of the bilinear equation of
the 2DTL equation) \cite{Hirota_book,OKMS}
\begin{equation}
\left(\frac{1}{a}D_s-1\right) \tau_n(k+1)\cdot \tau_n(k)
 +\tau_{n+1}(k+1)\tau_{n-1}(k)= 0\,.
\end{equation}

Applying a two-reduction condition $p_i=-q_i$, $i=1, \cdots, N$, which implies 
$\tau_{n-1}\Bumpeq  \tau_{n+1}$, we obtain
\begin{eqnarray}
  -\left(\frac{1}{a}D_s -1\right)f_{k+1}\cdot f_k =g_{k+1}g_k\,, \label{2DDTD1} \\
 -\left(\frac{1}{a}D_s -1\right)g_{k+1}\cdot g_k =f_{k+1}g_k\,, \label{2DDTD2}
\end{eqnarray}
by letting $\tau_0(k)=f_k$, $\tau_1(k)=g_k$.

Letting $u_k=g_k/f_k$, Eqs.(\ref{2DDTD1}) and (\ref{2DDTD2}) are
equivalent to
\begin{eqnarray}
   - \frac{1}{a} \left(\ln \frac{f_{k+1}}{f_k} \right)_{s}+1 &=& u_{k+1}u_k \,,\label{Dlog1} \\
  - \frac{1}{a} \left(\ln \frac{g_{k+1}}{g_k} \right)_{s}+1 &=& u^{-1}_{k+1}u^{-1}_k\,. \label{Dlog2}
\end{eqnarray}
Subtracting Eq.(\ref{Dlog2}) from Eq.(\ref{Dlog1}), one obtains
\begin{equation}\label{2DDTD2c}
      \frac{u_{k+1}u_k}{a} \left(\ln \frac{u_{k+1}}{u_k}
      \right)_{s} +1 = u^2_{k+1}u^2_k\,.
\end{equation}

Introducing the discrete analogue of hodograph transformation
$$
x_k=2ka-2(\ln g_k)_s\,,
$$
and
$$
\delta_k= x_{k+1}-x_k = 2a -2 \left(\ln \frac{g_{k+1}}{g_k} \right)_s \,.
$$
It then follows from Eq.(\ref{Dlog2})
$$
\delta_k= \frac{2a}{u_{k+1}u_k}\,,
$$
or
\begin{equation}\label{phodelta}
  \rho_{k+1} \rho_k = \frac{4a^2}{\delta_k^2}\,,
\end{equation}
by assuming $\rho_k = u^2_k$.

Introducing the dependent variable transformation
$$
w_k=-2 (\ln g_k)_{ss}\,,
$$
Eq.(\ref{2DDTD2c}) becomes
\begin{equation}\label{2DDTD2e}
\frac{1}{\delta_k} \left(\ln  \frac{\rho_{k+1}}{\rho_k}\right)_{s}+1
-\frac{4a^2}{\delta_k^2} =0\,.
\end{equation}
Differentiating Eq.(\ref{Dlog2}) with respect to $s$, we have
$$
\frac{1}{2a} (w_{k+1}-w_k) = -\frac{1}{u_{k+1}u_k}
\left( \ln u_{k+1}u_k \right)_{s}
=-\frac{1}{2u_{k+1}u_k} \left( \ln \rho_{k+1}\rho_k \right)_{s}\,,
$$
or
\begin{equation}\label{DBL1}
(\ln \rho_{k+1}\rho_k)_s = -\frac {2}{\delta_k} (w_{k+1}-w_k)\,.
\end{equation}
Eliminating $\rho_k$ and $\rho_{k+1}$ from
Eqs.(\ref{2DDTD2e}) and (\ref{DBL1}), we obtain
\begin{equation}\label{SemiSH1}
\fl \frac {1}{\delta_k} (w_{k+1}-w_k)- \frac {1}{\delta_{k-1}}
 (w_{k}-w_{k-1})
=\frac 12 (\delta_k +\delta_{k-1})-2a^2
\left( \frac {1}{\delta_k}+\frac {1}{\delta_{k-1}} \right)\,,
\end{equation}
or
\begin{equation}\label{SemiSH1b}
\Delta^2w_k= \frac {1}{\delta_k} M\left(\delta_k -\frac{4a^2}
{\delta_k} \right)\,,
\end{equation}
by defining a difference operator $\Delta$ and an average operator $M$ as follows
$$
\Delta F_k=\frac{F_{k+1}-F_k}{\delta_k}\,, \quad MF_k=\frac{F_{k+1}+F_k}{2}\,.
$$

Furthermore, a substitution of Eq.(\ref{phodelta}) into Eq.
(\ref{DBL1}) leads to
\begin{equation}\label{SemiSH2}
\frac{d \delta_k}{d s}  = w_{k+1}-w_k\,.
\end{equation}
Equations (\ref{SemiSH1}) and (\ref{SemiSH2}) constitute the
semi-discrete analogue of the SCHE.

Next, let us show that in the continuous limit, $a\to 0$
($\delta_k\to 0$), the proposed semi-discrete SCHE recovers
the continuous SCHE. To this end, Eqs.(\ref{SemiSH1}) and
(\ref{SemiSH2}) are rewritten as
$$
\left\{\begin{array}{l} \displaystyle
\frac{-2}{\delta_k+\delta_{k-1}} \left(\Delta w_{k}-\Delta
w_{k-1}\right) + 1 = \frac{4a^2}{\delta_k\delta_{k-1}} \,,
\\[15pt]\displaystyle
\partial_s\delta_k=w_{k+1}-w_k \,.
\end{array} \right.
$$
By taking logarithmic derivative of the first equation, we get
$$
\displaystyle
\frac{\displaystyle\partial_s\left\{\frac{-2}{\delta_k+\delta_{k-1}}
\left(\Delta w_{k}-\Delta w_{k-1}\right) +1\right\}}
{\displaystyle\frac{-2}{\delta_k+\delta_{k-1}} \left(\Delta
w_{k}-\Delta w_{k-1}\right) +1}
=-\frac{\partial_s\delta_k}{\delta_k}
-\frac{\partial_s\delta_{k-1}}{\delta_{k-1}}\,.
$$
The dependent variable $w$ is regarded as a function of $x$ and $t$,
where $x$ is the space coordinate of
the $k$-th lattice point and $t$ is the time, defined by
$$
x_k=x_0+\sum_{j=0}^{k-1}\delta_j\,,\qquad t=s\,.
$$
In the continuous limit, $a\to 0$ ($\delta_k\to 0$), we have
$$
\frac{\partial_s\delta_k}{\delta_k}=\frac{w_{k+1}-w_k}{\delta_k} \to w_x\,, \quad
\frac{\partial_s\delta_{k-1}}{\delta_{k-1}}=\frac{w_{k}-w_{k-1}}{\delta_{k-1}} \to w_x\,,
$$
$$
 \frac{2}{\delta_k+\delta_{k-1}}
\left(\Delta w_{k}-\Delta w_{k-1}\right) \to w_{xx}\,,
$$
$$
\frac{\partial x_k}{\partial s} =\frac{\partial x_0}{\partial s}
+\sum_{j=0}^{k-1}\frac{\partial\delta_j}{\partial s} =\frac{\partial
x_0}{\partial s} +\sum_{j=0}^{k-1}(w_{j+1}-w_j) \to w\,,
$$
$$
\partial_s=\partial_t+\frac{\partial x}{\partial s}\partial_x
\to\partial_t+w\partial_x\,,
$$
where the origin of space coordinate $x_0$ is taken so that
$\displaystyle\frac{\partial x_0}{\partial s}$ cancels $w_0$.
Thus the above semi-discrete SCHE converges to
$$
\frac{(\partial_t+w\partial_x)(-w_{xx}+1)} {\displaystyle
-w_{xx}+1}=-2w_x\,,
$$
or
\begin{equation}
(\partial_t+w\partial_x)w_{xx}
=2w_x\left({\displaystyle
-w_{xx}+1}\right)\,,
\end{equation}
which is nothing but the SCHE (\ref{HS-eq2}).

In summary, the semi-discrete analogue of the SCHE and its
determinant solution are given as follows:\\
{\bf The semi-discrete analogue of the SCHE}
\begin{equation}\label{semidiscreteSCHE}
\fl \left\{\begin{array}{l} \displaystyle
\frac {1}{\delta_k} (w_{k+1}-w_k)- \frac {1}{\delta_{k-1}}
 (w_{k}-w_{k-1})
=\frac 12 (\delta_k +\delta_{k-1})-2a^2
\left( \frac {1}{\delta_k}+\frac {1}{\delta_{k-1}} \right)
\,,
\\[15pt]\displaystyle
\frac{d \delta_k}{d t}  = w_{k+1}-w_k \,.
\end{array} \right.
\end{equation}
{\bf The determinant solution of the semi-discrete SCHE}
$$
w_k=-2 (\ln g_k)_{ss}\,,
$$
$$
\delta_k= x_{k+1}-x_k= 2a \frac{f_{k+1} f_k}{g_{k+1}g_k}\,,
$$
$$
\left\{\begin{array}{l}
x_k=2ka- 2(\ln g_k)_s\,,
\\
t=s\,,
\end{array}\right.
$$
$$
g_k=\left|
\psi_{i}^{(j)}(k)
\right|_{1\leq i,j\leq N}\,, \quad f_k=\left|
\psi_{i}^{(j-1)}(k)
\right|_{1\leq i,j\leq N}\,,
$$
\begin{equation}
\psi_i^{(j)}(k)=a_{i,1}p_i^j(1-ap_i)^{-k}e^{{p_i}^{-1} s+\eta_{0i}}
+a_{i,2}(-p_i)^j(1+ap_i)^{-k}e^{-{p_i}^{-1}s+\eta'_{0i}}\,.
\end{equation}

Introducing new independent variables $X_k=x_k/\kappa$ and $T=t/\kappa$,
we can include the parameter $\kappa$ in the semi-discrete SCHE
(\ref{semidiscreteSCHE})
\begin{equation}\label{semidiscreteSCHE2}
\fl \left\{\begin{array}{l} \displaystyle
\frac {1}{\delta_k} (w_{k+1}-w_k)- \frac {1}{\delta_{k-1}}
 (w_{k}-w_{k-1})
=\frac{1}{2\kappa^2} (\delta_k +\delta_{k-1})-2a^2 \left( \frac
{1}{\delta_k}+\frac {1}{\delta_{k-1}} \right) \,,
\\[15pt]\displaystyle
\frac{d \delta_k}{d T}  = w_{k+1}-w_k \,,
\end{array} \right.
\end{equation}
where $\delta_k=X_{k+1}-X_k$ and $s=\kappa T$.
This is the semi-discrete analogue of the SCHE (\ref{HS-eq}).

The $N$-cuspon solution of the semi-discrete SCHE
(\ref{semidiscreteSCHE2}) with the parameter $\kappa$ is given by
$$
w_k=-2(\ln g_k)_{ss}\,,
$$
$$
\delta_k=X_{k+1}-X_k= \frac{2a}{\kappa} \frac{f_{k+1} f_k}{g_{k+1}g_k}\,,
$$
$$ \displaystyle
\left\{\begin{array}{l}
X_k=\frac{2ka}{\kappa}-\frac{2}{\kappa}(\ln g_k)_{s}\,,
\\
T=\frac{s}{\kappa}\,,
\end{array}\right.
$$
$$
g_k=\left|
\psi_{i}^{(j)}(k)
\right|_{1\leq i,j\leq N}\,,\quad f_k=\left|
\psi_{i}^{(j-1)}(k)
\right|_{1\leq i,j\leq N}\,,
$$
\begin{equation}
\fl \psi_i^{(j)}(k)=a_{i,1}p_i^j(1-ap_i)^{-k}e^{{p_i}^{-1} s+\eta_{0i}}
+a_{i,2}(-p_i)^j(1+ap_i)^{-k}e^{-{p_i}^{-1}s+\eta'_{0i}}\,.
\end{equation}

\section{Full-discretization of the SCHE}
In much the same way of finding the semi-discrete
analogue of the SCHE, we seek for its full-discrete
analogue and in the process we arrive at its $N$-cuspon solution.

Consider the following Casorati determinant
\begin{equation}
\tau_n(k,l)=\left|\psi_i^{(n+j-1)}(k,l)\right|_{1\le i,j\le N}\,,
\end{equation}
where
\[
\fl \psi_i^{(n)}(k,l)
=a_{i,1}p_i^n(1-ap_i)^{-k}\left(1-b{p_i}^{-1}\right)^{-l}
e^{\xi_i}
+a_{i,2}q_i^n(1-aq_i)^{-k}\left(1-b{q_i}^{-1}\right)^{-l}
e^{\eta_i}\,,
\]
with
$$
\xi_i={p_i}^{-1}s+\xi_{i0}\,, \quad \eta_i={q_i}^{-1}s+\eta_{i0}\,.
$$
It is known that the above determinant satisfies
bilinear equations \cite{OKMS}
\begin{equation}
\left(\frac{1}{a}D_s-1\right)\tau_n(k+1,l)\cdot\tau_n(k,l)
+\tau_{n+1}(k+1,l)\tau_{n-1}(k,l)=0\,,
\end{equation}
and
\begin{equation}
(bD_s-1)\tau_{n}(k,l+1)\cdot\tau_{n+1}(k,l)
+\tau_{n}(k,l)\tau_{n+1}(k,l+1)=0\,.
\end{equation}
Here $a, b$ are mesh sizes for space and time variables,
respectively.

Applying the two-reduction $\tau_{n-1}=(\prod_{i=1}^Np^2_i)^{-1}
\tau_{n+1}$, i.e., enforcing $p_i=-q_i$, $i=1, \cdots, N$,
and letting $\tau_0(k,l)=f_{k,l}$, $\tau_1(k,l)=g_{k,l}$,
the above bilinear equations take the following form:
\begin{eqnarray}
&&\left(\frac{1}{a}D_s-1\right)f_{k+1,l}\cdot f_{k,l}
+{g}_{k+1,l}{g}_{k,l}=0\,,\\
&&\left(\frac{1}{a}D_s-1\right){g}_{k+1,l}\cdot {g}_{k,l}
+{f}_{k+1,l}{f}_{k,l}=0\,,\\
&&(bD_s-1)f_{k,l+1}\cdot {g}_{k,l}
+f_{k,l}{g}_{k,l+1}=0\,,\\
&&(bD_s-1){g}_{k,l+1}\cdot f_{k,l}
+{g}_{k,l}f_{k,l+1}=0\,,
\end{eqnarray}
where the gauge transformation $\tau_n \to (\prod_{i=1}^Np_i)^{n}
\tau_n$ is used. It is readily shown that the above equations are
equivalent to
\begin{eqnarray}
&&\frac{1}{a}\left(\ln\frac{f_{k+1,l}}{f_{k,l}}\right)_s=1
-\frac{{g}_{k+1,l}{g}_{k,l}}{f_{k+1,l}f_{k,l}}\,, \label{fd_log1}\\
&&\frac{1}{a}\left(\ln\frac{{g}_{k+1,l}}{{g}_{k,l}}\right)_s=1
-\frac{f_{k+1,l}f_{k,l}}{{g}_{k+1,l}{g}_{k,l}}\,, \label{fd_log2}\\
&&b\left(\ln\frac{f_{k,l+1}}{{g}_{k,l}}\right)_s=1
-\frac{f_{k,l}{g}_{k,l+1}}{f_{k,l+1}{g}_{k,l}}\,, \label{fd_log3}\\
&&b\left(\ln\frac{{g}_{k,l+1}}{f_{k,l}}\right)_s=1
-\frac{{g}_{k,l}f_{k,l+1}}{{g}_{k,l+1}f_{k,l}}\,. \label{fd_log4}
\end{eqnarray}
We introduce a dependent variable transformation
\begin{equation}\label{w-g}
w_{k,l}=-2\left(\ln{g_{k,l}}\right)_{ss}\,,
\end{equation} 
and a discrete hodograph transformation 
\begin{equation}\label{g}
x_{k,l}=2ka-2(\ln g_{k,l})_s\,,
\end{equation}
then the mesh  
\begin{equation}
\delta_{k,l}=x_{k+1,l}-x_{k,l}=2a-2\left(\ln{\frac{g_{k+1,l}}{g_{k,l}}}\right)_{s}\, 
\end{equation}
is naturally defined. It then follows
\begin{equation}
\left(\ln{\frac{g_{k+1,l}}{g_{k-1,l}}}\right)_{s}=2a-\frac 12\left(\delta_{k,l}+ \delta_{k-1,l}\right)\,.
\label{fd-logg}
\end{equation}
In view of Eq.(\ref{fd_log2}), one obtains
\begin{equation}
\frac{f_{k+1,l}f_{k,l}}{{g}_{k+1,l}{g}_{k,l}} = \frac{\delta_{k,l}}{2a}\,.
\end{equation}
A substitution into Eq.(\ref{fd_log1}) yields
\begin{equation}
\left(\ln{\frac{f_{k+1,l}}{f_{k,l}}}\right)_{s}=a-\frac {2a^2}{\delta_{k,l}}\,,
\label{fd-logf1}
\end{equation}
it then follows
\begin{equation}
\left(\ln{\frac{f_{k+1,l}}{f_{k-1,l}}}\right)_{s}
=2a-{2a^2}\left(\frac{1}{\delta_{k,l}}+\frac{1}{\delta_{k-1,l}}\right)\,.
\label{fd-logf}
\end{equation}
Starting from an alternative form of Eq.(\ref{fd_log2})
\begin{equation}
2a-2\left(\ln\frac{{g}_{k+1,l}}{{g}_{k,l}}\right)_s
=2a \frac{f_{k+1,l}f_{k,l}}{{g}_{k+1,l}{g}_{k,l}}\,,
\end{equation}
we obtain
\begin{equation}
\frac{w_{k+1,l}-w_{k,l}}{\delta_{k,l}}=
\frac{-2\left(\ln\frac{{g}_{k+1,l}}{{g}_{k,l}}\right)_{ss}}
{2a-2\left(\ln\frac{{g}_{k+1,l}}{{g}_{k,l}}\right)_s}=
\left( \ln \frac{f_{k+1,l}f_{k,l}}{{g}_{k+1,l}{g}_{k,l}}\right)_s \,,
\label{fd-w}
\end{equation}
by taking logarithmic derivative with respect to $s$.
A shift from $k$ to $k-1$ gives
\begin{equation}
\frac{w_{k,l}-w_{k-1,l}}{\delta_{k-1,l}}=
\left( \ln \frac{f_{k,l}f_{k-1,l}}{{g}_{k,l}{g}_{k-1,l}}\right)_s \,.
\label{fd-w2}
\end{equation}
Subtracting Eq.(\ref{fd-w2}) from Eq.(\ref{fd-w}), we obtain
\begin{equation}
\frac{w_{k+1,l}-w_{k,l}}{\delta_{k,l}}-\frac{w_{k,l}-w_{k-1,l}}{\delta_{k-1,l}}=
\left(\ln{\frac{f_{k+1,l}}{f_{k-1,l}}}\right)_{s} - \left(\ln{\frac{g_{k+1,l}}{g_{k-1,l}}}\right)_{s}\,.
\end{equation}
By using the relations (\ref{fd-logg}) and (\ref{fd-logf}), we
finally arrive at
\begin{equation}
\fl\frac{w_{k+1,l}-w_{k,l}}{\delta_{k,l}}-\frac{w_{k,l}-w_{k-1,l}}{\delta_{k-1,l}}
-\frac 12\left(\delta_{k,l}+
\delta_{k-1,l}\right)+{2a^2}\left(\frac{1}{\delta_{k,l}}+\frac{1}{\delta_{k-1,l}}\right)
 =0\,.
 \label{fd-sch1}
\end{equation}
Similar to Eq.(\ref{SemiSH1b}), Eq.(\ref{fd-sch1}) constitutes the
first equation of the full-discretization of the SCHE, which
can be cast into a simpler form:
\begin{equation}\label{fullSCH1}
\Delta^2w_{k,l}= \frac {1}{\delta_{k,l}} M\left(\delta_{k,l} -\frac{4a^2}
{\delta_{k,l}} \right)\,.
\end{equation}

Next, we seek for the second equation of the full-discretization.
Recalling (\ref{fd_log1})--(\ref{fd_log4}), one could obtain
\begin{equation}
\frac{x_{k+1,l+1}-x_{k,l+1}}{x_{k+1,l}-x_{k,l}}=
\frac{2a-2\left(\ln\frac{{g}_{k+1,l+1}}{{g}_{k,l+1}}\right)_s}
{2a-2\left(\ln\frac{{g}_{k+1,l}}{{g}_{k,l}}\right)_s}
=\frac{\left(\ln\frac{{g}_{k+1,l+1}}{{f}_{k+1,l}}\right)_s-
\frac{1}{b}}{\left(\ln\frac{{f}_{k,l+1}}{{g}_{k,l}}\right)_s-\frac{1}{b}}\,,
\label{fd-2key}
\end{equation}
here a shift from $l$ to $l+1$ in (\ref{fd_log2}) and a shift from $k$
to $k+1$ in (\ref{fd_log4}) are employed.

From Eqs.(\ref{w-g}), (\ref{fd-logf1}) and (\ref{fd-w}),
one can find the following
two relations
\begin{equation}
\fl \left(\ln\frac{{g}_{k+1,l+1}}{{f}_{k+1,l}}\right)_s = - \frac{w_{k+1,l}-w_{k,l}-2a^2}{2 \delta_{k,l}}
+ \frac 14 \left(x_{k+1,l}+x_{k,l}-2 x_{k+1,l+1} \right)\,,
\end{equation}
\begin{equation}
\fl \left(\ln\frac{{f}_{k,l+1}}{{g}_{k,l}}\right)_s =  \frac{w_{k+1,l+1}-w_{k,l+1}+2a^2}{2 \delta_{k,l+1}}
-\frac 14 \left(x_{k+1,l+1}+x_{k,l+1}-2 x_{k,l} \right)\,,
\end{equation}
after some tedious algebraic manipulations. Substituting these two
relations into (\ref{fd-2key}), we finally obtain the second
equation of the fully discrete analogue of the SCHE
\begin{eqnarray}
   \frac{\delta_{k,l+1}- \delta_{k,l}} {b} &+& \frac 14 \delta_{k,l+1}
 \left(x_{k+1,l+1}+x_{k,l+1}-2 x_{k,l} \right) \nonumber \\
 & + & \frac 14 \delta_{k,l}
 \left(x_{k+1,l}+x_{k,l}-2 x_{k+1,l+1}  \right) \nonumber \\
 &=& \frac 12 \left( w_{k+1,l+1}+w_{k+1,l}-
 w_{k,l+1}-w_{k,l}\right)\,.
\end{eqnarray}
Taking the continuous limit $b\to 0$ in time, we have
$$
\frac{\delta_{k,l+1}- \delta_{k,l}} {b} \to \frac{d \delta_k}{d s}\,,
$$
$$
\delta_{k,l+1}
 \left(x_{k+1,l+1}+x_{k,l+1}-2 x_{k,l} \right) \to 0\,,
$$
$$
\delta_{k,l+1}
 \delta_{k,l}
 \left(x_{k+1,l}+x_{k,l}-2 x_{k+1,l+1}  \right) \to 0\,,
$$
and
$$
\frac 12 \left( w_{k+1,l+1}+w_{k+1,l}-
 w_{k,l+1}-w_{k,l}\right) \to w_{k+1}-w_k.
$$
Therefore, one recovers exactly
the second equation of the semi-discrete SCHE
(\ref{SemiSH2}).

In summary, the fully discrete analogue of the SCHE and its determinant
solution are given as follows:\\
{\bf The fully discrete analogue of the SCHE}
\begin{equation}\label{fulldiscreteSCHE}
\fl \left\{\begin{array}{l} \displaystyle
\frac{w_{k+1,l}-w_{k,l}}{\delta_{k,l}}-\frac{w_{k,l}-w_{k-1,l}}{\delta_{k-1,l}}
-\frac 12\left(\delta_{k,l}+
\delta_{k-1,l}\right)+{2a^2}\left(\frac{1}{\delta_{k,l}}+\frac{1}{\delta_{k-1,l}}\right)
 =0\,,
\\[15pt]\displaystyle
 \frac{\delta_{k,l+1}- \delta_{k,l}} {b}
+ \frac 14 \delta_{k,l+1}
 \left(x_{k+1,l+1}+x_{k,l+1}-2 x_{k,l} \right)
\\[15pt]\displaystyle
\qquad  +  \frac 14 \delta_{k,l}
 \left(x_{k+1,l}+x_{k,l}-2 x_{k+1,l+1}  \right)
= \frac 12 \left( w_{k+1,l+1}+w_{k+1,l}-
 w_{k,l+1}-w_{k,l}\right)\,.
\end{array} \right.
\end{equation}
{\bf The determinant solution of the fully discrete SCHE}
$$
w_{k,l}=-2 (\ln g_{k,l})_{ss}=-2\frac{{\bar h_{k,l}}g_{k,l}-h_{k,l}^2}{g_{k,l}^2}\,,
$$

$$
x_{k,l}=2ka- 2(\ln g_{k,l})_s=2ka-2\frac{h_{k,l}}{g_{k,l}}\,,   
$$                                                              
$$
\delta_{k,l}= x_{k+1,l}-x_{k,l}= 2a \frac{f_{k+1,l} f_{k,l}}{g_{k+1,l}g_{k,l}}\,,
$$
$$
g_{k,l}=\left|
\psi_{i}^{(j)}(k,l)
\right|_{1\leq i,j\leq N}\,, \quad f_{k,l}=\left|
\psi_{i}^{(j-1)}(k,l)
\right|_{1\leq i,j\leq N}\,,
$$
\begin{eqnarray*}
&&h_{k,l}=\frac{\partial g_{k,l}}{\partial s}=
\left|\matrix{
\psi_1^{(0)}(k,l) &\psi_1^{(2)}(k,l) &\psi_1^{(3)}(k,l) &\cdots &\psi_1^{(N)}(k,l) \cr
\psi_2^{(0)}(k,l) &\psi_2^{(2)}(k,l) &\psi_2^{(3)}(k,l) &\cdots &\psi_2^{(N)}(k,l) \cr
\vdots            &\vdots             &\vdots & \ddots       &\vdots                \cr
\psi_N^{(0)}(k,l) &\psi_N^{(2)}(k,l) &\psi_N^{(3)}(k,l) &\cdots
 &\psi_N^{(N)}(k,l)}\right|
\,,\\
&&{\bar h}_{k,l}=\frac{\partial^2 g_{k,l}}{\partial s^2}=
\left|\matrix{
\psi_1^{(-1)}(k,l) &\psi_1^{(2)}(k,l) &\psi_1^{(3)}(k,l) &\cdots &\psi_1^{(N)}(k,l) \cr
\psi_2^{(-1)}(k,l) &\psi_2^{(2)}(k,l) &\psi_2^{(3)}(k,l) &\cdots &\psi_2^{(N)}(k,l) \cr
\vdots            &\vdots             &\vdots &  \ddots     &\vdots                \cr
\psi_N^{(-1)}(k,l) &\psi_N^{(2)}(k,l) &\psi_N^{(3)}(k,l) &\cdots
 &\psi_N^{(N)}(k,l)}\right|
\\
&&\qquad +
\left|\matrix{
\psi_1^{(0)}(k,l) &\psi_1^{(1)}(k,l) &\psi_1^{(3)}(k,l) &\cdots &\psi_1^{(N)}(k,l) \cr
\psi_2^{(0)}(k,l) &\psi_2^{(1)}(k,l) &\psi_2^{(3)}(k,l) &\cdots &\psi_2^{(N)}(k,l) \cr
\vdots            &\vdots            &\vdots  &  \ddots     &\vdots                \cr
\psi_N^{(0)}(k,l) &\psi_N^{(1)}(k,l) &\psi_N^{(3)}(k,l) &\cdots
 &\psi_N^{(N)}(k,l)}\right|
\,,
\end{eqnarray*}
$$
\psi_i^{(j)}(k,l)=a_{i,1}p_i^j(1-ap_i)^{-k}
\left(1-b{p_i}^{-1}\right)^{-l}
e^{\xi_i}
+a_{i,2}(-p_i)^j(1+ap_i)^{-k}\left(1+b{p_i}^{-1}\right)^{-l}e^{\eta_i}\,,
$$
\begin{equation}
\xi_i={p_i}^{-1}s+\xi_{i0}\,, \quad \eta_i=-{p_i}^{-1}s+\eta_{i0}\,.
\end{equation}
Note that $s$ is an auxiliary parameter.
By virtue of $s$,
$h_{k,l}$ and ${\bar h}_{k,l}$
can be expressed as $h_{k,l} = \partial_s g_{k,l}$ and
${\bar h}_{k,l} = \partial^2_s g_{k,l}$, respectively,
because the auxiliary parameter
$s$ works on elements of the above determinant
by $\partial_s \psi_i^{(n)}(k,l) = \psi_i^{(n-1)}(k,l)$.

Introducing new independent variables $X_{k,l}=x_{k,l}/\kappa$ and $\tilde{b}=b/\kappa$,
we can include the parameter $\kappa$ in the full-discrete SCHE
(\ref{fulldiscreteSCHE}):
\begin{equation}\label{fulldiscreteSCHE2}
\fl \left\{\begin{array}{l} \displaystyle
\frac{w_{k+1,l}-w_{k,l}}{\delta_{k,l}}-\frac{w_{k,l}-w_{k-1,l}}{\delta_{k-1,l}}
-\frac {1}{2\kappa^2}\left(\delta_{k,l}+
\delta_{k-1,l}\right)+{2a^2}\left(\frac{1}{\delta_{k,l}}+\frac{1}{\delta_{k-1,l}}\right)
 =0\,,
\\[15pt]\displaystyle
 \frac{\delta_{k,l+1}- \delta_{k,l}} {\tilde{b}}
+ \frac {1}{4\kappa^2} \delta_{k,l+1}
 \left(X_{k+1,l+1}+X_{k,l+1}-2 X_{k,l} \right)
\\[15pt]\displaystyle
\qquad  +  \frac {1}{4\kappa^2} \delta_{k,l}
 \left(X_{k+1,l}+X_{k,l}-2 X_{k+1,l+1}  \right)
= \frac 12 \left( w_{k+1,l+1}+w_{k+1,l}-
 w_{k,l+1}-w_{k,l}\right)\,.
\end{array} \right.
\end{equation}

Similarly, the $N$-cuspon solution of the full-discrete SCHE
(\ref{fulldiscreteSCHE2}) with the parameter $\kappa$ is given as follows:
$$
w_{k,l}=-2(\ln g_{k,l})_{ss}=-{2}
\frac{{\bar h_{k,l}}g_{k,l}-h_{k,l}^2}{g_{k,l}^2}\,,
$$

$$
X_{k,l}=\frac{2ka}{\kappa}-\frac{2}{\kappa}(\ln g_{k,l})_{s}
=\frac{2ka}{\kappa}-\frac{2}{\kappa}\frac{h_{k,l}}{g_{k,l}}\,,
$$
$$
\delta_{k,l}=X_{k+1,l}-X_{k,l}= \frac{2a}{\kappa} \frac{f_{k+1,l} f_{k,l}}{g_{k+1,l}g_{k,l}}\,,
$$

$$
g_{k,l}=\left|
\psi_{i}^{(j)}(k,l)
\right|_{1\leq i,j\leq N}\,, \quad f_{k,l}=\left|
\psi_{i}^{(j-1)}(k,l)
\right|_{1\leq i,j\leq N}\,,
$$
\begin{eqnarray*}
&&h_{k,l}=\frac{\partial g_{k,l}}{\partial s}=
\frac{1}{\kappa}\left|\matrix{
\psi_1^{(0)}(k,l) &\psi_1^{(2)}(k,l) &\psi_1^{(3)}(k,l) &\cdots &\psi_1^{(N)}(k,l) \cr
\psi_2^{(0)}(k,l) &\psi_2^{(2)}(k,l) &\psi_2^{(3)}(k,l) &\cdots &\psi_2^{(N)}(k,l) \cr
\vdots            &\vdots             &\vdots & \ddots       &\vdots                \cr
\psi_N^{(0)}(k,l) &\psi_N^{(2)}(k,l) &\psi_N^{(3)}(k,l) &\cdots
 &\psi_N^{(N)}(k,l)}\right|
\,,\\
&&{\bar h}_{k,l}=\frac{\partial^2 g_{k,l}}{\partial s^2}=
\frac{1}{\kappa^2}\left|\matrix{
\psi_1^{(-1)}(k,l) &\psi_1^{(2)}(k,l) &\psi_1^{(3)}(k,l) &\cdots &\psi_1^{(N)}(k,l) \cr
\psi_2^{(-1)}(k,l) &\psi_2^{(2)}(k,l) &\psi_2^{(3)}(k,l) &\cdots &\psi_2^{(N)}(k,l) \cr
\vdots            &\vdots             &\vdots &  \ddots     &\vdots                \cr
\psi_N^{(-1)}(k,l) &\psi_N^{(2)}(k,l) &\psi_N^{(3)}(k,l) &\cdots
 &\psi_N^{(N)}(k,l)}\right|
\\
&&\qquad +
\frac{1}{\kappa^2}\left|\matrix{
\psi_1^{(0)}(k,l) &\psi_1^{(1)}(k,l) &\psi_1^{(3)}(k,l) &\cdots &\psi_1^{(N)}(k,l) \cr
\psi_2^{(0)}(k,l) &\psi_2^{(1)}(k,l) &\psi_2^{(3)}(k,l) &\cdots &\psi_2^{(N)}(k,l) \cr
\vdots            &\vdots            &\vdots  &  \ddots     &\vdots                \cr
\psi_N^{(0)}(k,l) &\psi_N^{(1)}(k,l) &\psi_N^{(3)}(k,l) &\cdots
 &\psi_N^{(N)}(k,l)}\right|
\,,
\end{eqnarray*}
$$
\psi_i^{(j)}(k,l)=a_{i,1}p_i^j(1-ap_i)^{-k}
\left(1-b{p_i}^{-1}\right)^{-l}
e^{\xi_i}
+a_{i,2}(-p_i)^j(1+ap_i)^{-k}\left(1+b{p_i}^{-1}\right)^{-l}e^{\eta_i}\,,
$$
\begin{equation}
\xi_i={p_i}^{-1}s+\xi_{i0}\,, \quad \eta_i=-{p_i}^{-1}s+\eta_{i0}\,.
\end{equation}

\section{Concluding remarks}
In the present paper, bilinear equations and the determinant solution of
the SCHE are obtained from the two-reduction of 2DTL equations.
Based on this fact, integrable semi-and full-discrete analogues of 
the SCHE are constructed. The $N$-soliton solutions of both continuous and discrete SCHEs
are formulated in the form of the Casorati determinant.
Note that the short pulse equation was also obtained from the
two-reduction of the 2DTL equation \cite{FMO09}.

Finally, we remark that the present paper is one of our series of
work in an attempt of obtaining integrable discrete analogues for a
class of integrable nonlienar PDEs whose solutions possess
singularities such as peakon, cuspon or loop soliton solutions.
New discrete integrable
systems obtained in this paper, along with the semi-discrete analogue for the
Camassa-Holm equation \cite{OMF_JPA} and the semi-discrete and fully
discrete analogues of the short pulse equation
\cite{FMO09}
deserves further study in the future.

\end{document}